\documentclass[sigconf]{acmart}

\AtBeginDocument{%
  }

\setcopyright{acmlicensed}
\copyrightyear{2025}
\acmYear{2025}
\acmDOI{XXXXXXX.XXXXXXX}
\acmConference[MMAsia '25]{Proceedings of the 2025 ACM Multimedia Asia Conference}{December 09--12, 2025}{Kuala Lumpur, Malaysia}
\acmISBN{978-1-4503-XXXX-X/2018/06}

\begin{document}

\title{ECTSpeech: Enhancing Efficient Speech Synthesis via Easy Consistency Tuning}


\author{Tao Zhu\textsuperscript{1,2}}
\affiliation{%
  \institution{\textsuperscript{1}School of Computer Science and Technology, Xinjiang University; \\ \textsuperscript{2}Joint International Research Laboratory of Silk Road Multilingual Cognitive Computing}
  \city{Urumqi}
  \state{Xinjiang}
  \country{China}}
\email{107552304226@stu.xju.edu.cn}

\author{Yinfeng Yu\textsuperscript{1,2,}}
\authornote{Yinfeng Yu is the corresponding author.}
\affiliation{%
  \institution{\textsuperscript{1}School of Computer Science and Technology, Xinjiang University; \\ \textsuperscript{2}Joint International Research Laboratory of Silk Road Multilingual Cognitive Computing}
  \city{Urumqi}
  \state{Xinjiang}
  \country{China}}
\email{yuyinfeng@xju.edu.cn}

\author{Liejun Wang\textsuperscript{1,2}}
\affiliation{%
  \institution{\textsuperscript{1}School of Computer Science and Technology, Xinjiang University; \\
  \textsuperscript{2}Joint International Research Laboratory of Silk Road Multilingual Cognitive Computing}
  \city{Urumqi}
  \state{Xinjiang}
  \country{China}}
\email{wljxju@xju.edu.cn}

\author{Fuchun Sun\textsuperscript{3}}
\affiliation{%
  \institution{\textsuperscript{3}Tsinghua University}
  \city{Haidian}
  \state{Beijing}
  \country{China}}
\email{fcsun@mail.tsinghua.edu.cn}

\author{Wendong Zheng\textsuperscript{4}}
\affiliation{%
  \institution{\textsuperscript{4}Tianjin University of Technology}
  \city{Tianjin}
  \country{China}}
\email{zwendong@email.tjut.edu.cn}

\renewcommand{\shortauthors}{T.Z, Y.Y.F et al.}

\begin{abstract}
Diffusion models have demonstrated remarkable performance in speech synthesis, but typically require multi-step sampling, resulting in low inference efficiency. Recent studies address this issue by distilling diffusion models into consistency models, enabling efficient one-step generation. However, these approaches introduce additional training costs and rely heavily on the performance of pre-trained teacher models. In this paper, we propose ECTSpeech, a simple and effective one-step speech synthesis framework that, for the first time, incorporates the Easy Consistency Tuning (ECT) strategy into speech synthesis. By progressively tightening consistency constraints on a pre-trained diffusion model, ECTSpeech achieves high-quality one-step generation while significantly reducing training complexity. In addition, we design a multi-scale gate module (MSGate) to enhance the denoiser's ability to fuse features at different scales. Experimental results on the LJSpeech dataset demonstrate that ECTSpeech achieves audio quality comparable to state-of-the-art methods under single-step sampling, while substantially reducing the model's training cost and complexity.
\end{abstract}





%%\received{20 February 2007}
%%\received[revised]{12 March 2009}
%%\received[accepted]{5 June 2009}

\maketitle

\section{Introduction}

\begin{figure}
    \centering
    \includegraphics[width=1\linewidth]{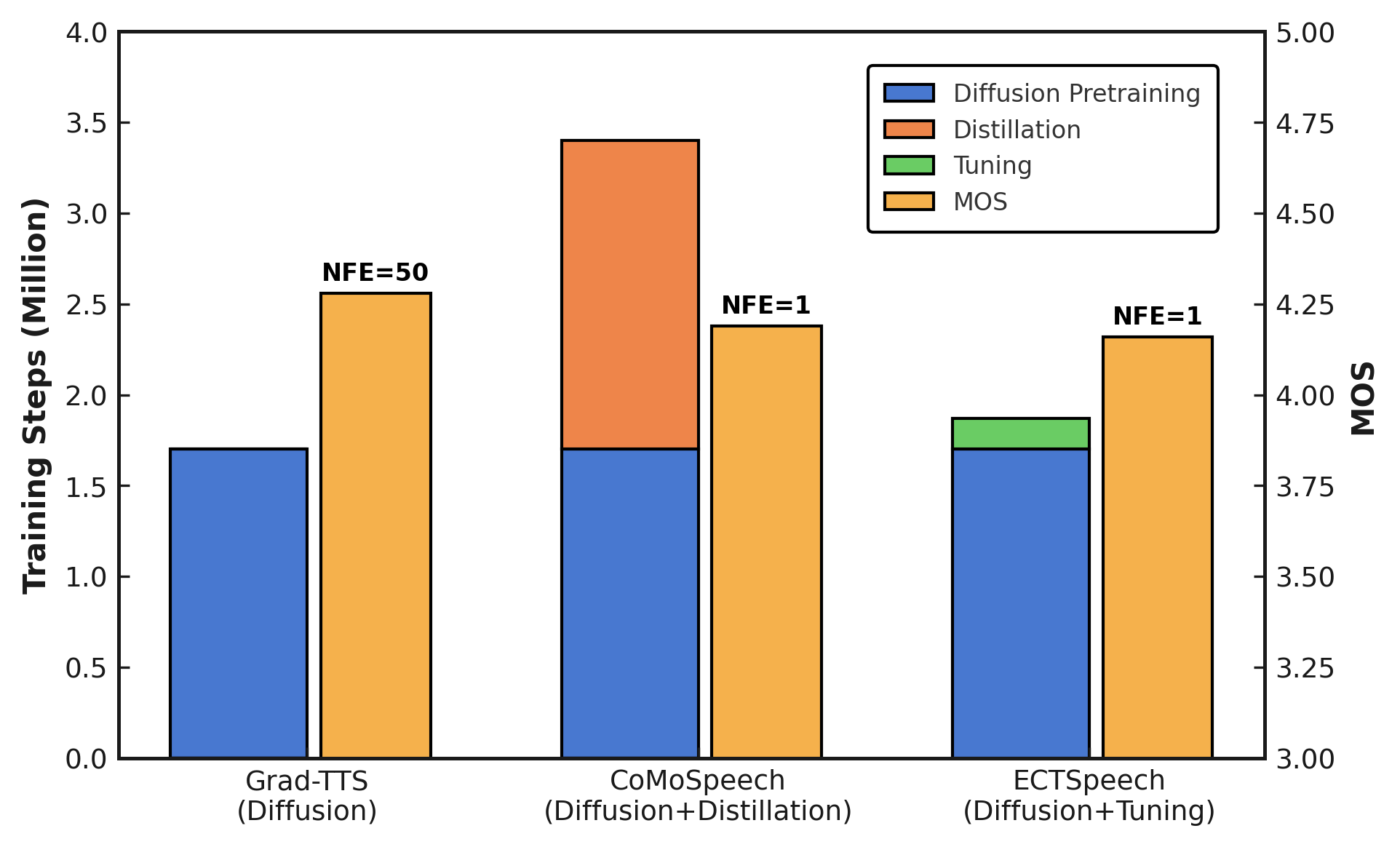}
    \caption{Training cost and synthesis quality of different methods on the LJSpeech dataset. The horizontal axis denotes different methods (Grad-TTS, CoMoSpeech, ECTSpeech) and their training pipelines. The left vertical axis indicates the number of training steps (in millions), and the right vertical axis shows MOS scores. NFE represents the number of sampling steps used for inference in each method.}
    \label{fig:head figure}
\end{figure}

Text-to-Speech (TTS) aims to convert textual information into natural and intelligible speech, playing a crucial role in applications such as intelligent voice assistants, content broadcasting, and virtual humans. With the continuous development of deep neural network methods, modern TTS systems have achieved significant improvements in naturalness, expressiveness, and flexibility\cite{yu2021weavenet,jiao2024mfhca}. Modern neural TTS systems typically adopt a two-stage architecture, where an acoustic model converts text into Mel-spectrograms and a vocoder reconstructs waveforms. Since acoustic features largely determine the naturalness of synthesized speech, recent research has mainly focused on improving the performance of the acoustic modeling stage.

In recent years, representative models such as the Tacotron series~\cite{Tacotron,Tacotron2}, Transformer-TTS~\cite{Transformer-TTS}, and the FastSpeech series~\cite{fastspeech,fastspeech2} have significantly advanced both the quality and inference efficiency of speech synthesis. However, there is still room for further improvement in acoustic modeling. As a result, diffusion models capable of generating high-quality samples have attracted increasing attention in the field of speech synthesis.

Diffusion models have become an important approach in the field of speech synthesis by gradually adding noise to data and learning to reverse this process through denoising. In TTS tasks, most diffusion-based methods require multi-step sampling to generate high-quality speech, gradually recovering the natural speech signal from noise. However, such methods~\cite{diffsinger,Grad-TTS,diff-tts} typically require a large number of sampling steps for each generation, resulting in slow overall inference speed. To address this issue, several acceleration schemes have been proposed. ResGrad~\cite{resgrad} estimates the residual between a pre-trained FastSpeech2~\cite{fastspeech2} model and real speech using a diffusion model. ProDiff~\cite{prodiff} improves inference efficiency by reducing the number of sampling steps through progressive distillation. DiffGAN-TTS~\cite{diffgan} approximates the denoising function via adversarial training, enabling efficient speech synthesis. In addition, CoMoSpeech~\cite{comospeech} adopts the consistency model~\cite{CM2023} framework, distilling a student model from a pre-trained teacher diffusion model to enable one-step high-quality speech generation.

Despite the improvements in inference efficiency achieved by the above methods, they often incur additional training costs or require complex distillation pipelines. For example, CoMoSpeech~\cite{comospeech} adopts the Consistency Model (CM)~\cite{CM2023} framework, where a teacher diffusion model is distilled into a student model capable of one-step sampling, thus enabling efficient speech synthesis. However, consistency distillation requires training two separate models and is highly dependent on the performance of the teacher, leading to increased training complexity. Recently, the Easy Consistency Tuning (ECT)~\cite{geng2025consistency} strategy proposed in the image generation domain enables efficient and stable one-step generation by progressively tightening consistency constraints on a pretrained diffusion model, but has not yet been applied to speech synthesis.

To address these challenges, we propose an efficient speech synthesis framework, ECTSpeech, which introduces the ECT strategy into the TTS domain for the first time. After pretraining a base diffusion acoustic model, we directly perform fine-tuning by progressively tightening the consistency constraint, enabling efficient one-step speech generation. Compared to CoMoSpeech, ECTSpeech does not require a separate student model, making the overall pipeline streamlined. Additionally, we introduce a multi-scale gate module (MSGate) to enhance the fusion of multi-scale features within the denoising network. Figure~\ref{fig:head figure} compares pure diffusion modeling with Grad-TTS, consistency distillation with CoMoSpeech, and our ECTSpeech in terms of training cost, synthesized speech quality, and the number of inference steps. It can be observed that ECTSpeech substantially reduces training cost while achieving speech quality comparable to or surpassing state-of-the-art methods. The contributions of our work are as follows:
\begin{itemize}
    \item[(1)] We introduce the Easy Consistency Tuning strategy into speech synthesis for the first time, enabling high-quality one-step speech generation without additional student model distillation;
    \item[(2)] We propose a novel MSGate to enhance the adaptive fusion of multi-scale information in the denoising network, effectively improving one-step speech synthesis quality;
    \item[(3)] Extensive experiments on the LJSpeech dataset demonstrate that our approach significantly reduces training cost and improves inference efficiency, while achieving speech naturalness comparable to or even surpassing current state-of-the-art methods.
\end{itemize}

\section{Background}

Diffusion Models (DMs) generate high-dimensional data by progressively adding Gaussian noise to the data distribution $p_{\mathrm{data}}(\mathbf{x}_0)$, constructing a noisy data sequence $\{\mathbf{x}_t\}_{t \in [0, T]}$, where $p_0(\mathbf{x}) = p_{\mathrm{data}}(\mathbf{x})$ and $p_T(\mathbf{x})$ converges to a Gaussian distribution. The original samples can then be reconstructed by a reverse denoising process, thus enabling generative modeling of complex data distributions~\cite{ho2020denoising,song2021scorebased}. The forward process can be described by a stochastic differential equation (SDE):
\begin{equation}
d\mathbf{x}_t = \mathbf{f}(\mathbf{x}_t, t)\,dt + g(t)\,d\mathbf{w}_t,
\end{equation}
where $\mathbf{w}_t$ denotes a standard Wiener process (Brownian motion), and $\mathbf{f}(\cdot, \cdot)$ and $g(\cdot)$ are the drift and diffusion coefficients, respectively.

The reverse process is equivalent to a Probability Flow Ordinary Differential Equation (PF-ODE)~\cite{karras2022elucidating}:
\begin{equation}
d\mathbf{x}_t = \left[ \mathbf{f}(\mathbf{x}_t, t) - \frac{1}{2}g^2(t)\nabla_{\mathbf{x}_t} \log p_t(\mathbf{x}_t) \right] dt,
\end{equation}
where $\nabla_{\mathbf{x}_t} \log p_t(\mathbf{x}_t)$ is the score function, representing the gradient of the data distribution at time $t$.

In practical sampling, the score function can be approximated by a neural network $f(\mathbf{x}_t, t)$ through minimizing the denoising error $\|f(\mathbf{x}_t, t) - \mathbf{x}_0\|^2$. That is,
\begin{equation}
\nabla_{\mathbf{x}_t} \log p_t(\mathbf{x}_t) = \frac{f(\mathbf{x}_t, t) - \mathbf{x}_t}{\sigma^2(t)},
\end{equation}
where $f(\mathbf{x}_t, t)$ denotes the denoiser network that takes the noisy sample $\mathbf{x}_t$ at time step $t$ as input and predicts the clean data $\mathbf{x}_0$, i.e., recovers $\mathbf{x}_t$ via denoising. With the combination of PF-ODE and neural network approximation, deterministic sampling is achievable, but inference typically requires multiple iterative steps.

To accelerate sampling, Consistency Models (CMs) were proposed~\cite{CM2023}. The core idea is to ensure that the network output remains consistent at any time point along the same PF-ODE sampling trajectory, that is,
\begin{equation}
f(\mathbf{x}_t, t) = f(\mathbf{x}_{t'}, t') = \mathbf{x}_0, \quad \forall\, t, t' \in [\epsilon, T],
\end{equation}
where the boundary condition $f(\mathbf{x}_0, 0) = \mathbf{x}_0$ guarantees that the output equals the input when there is no noise. This constraint ensures that any noisy sample on the sampling trajectory can be directly restored to the original data, enabling one-step generation.

Training strategies for consistency models mainly include consistency distillation from a pretrained diffusion model and consistency training~\cite{CM2023}. The former relies on a teacher model, while the latter does not require a teacher. Both approaches enforce the consistency of network outputs at different time steps through a consistency loss.

To further improve the training efficiency of consistency models, ECT~\cite{geng2025consistency} was proposed. ECT first conducts diffusion model pretraining, and then gradually tightens the consistency constraint. The training objective is defined as
\begin{equation}
\mathcal{L}_{\mathrm{ECT}} = \mathbb{E}_{\mathbf{x}_0, \boldsymbol{\epsilon}, t, r}\left[ w(t, r)\, d\left(f(\mathbf{x}_t, t), f_{\mathrm{sg}}(\mathbf{x}_r, r)\right)\right],
\end{equation}
where $\mathbf{x}_t = \mathbf{x}_0 + t\boldsymbol{\epsilon}$, $\mathbf{x}_r = \mathbf{x}_0 + r\boldsymbol{\epsilon}$ ($\boldsymbol{\epsilon}$ is Gaussian noise), $w(t, r)$ is a weighting function, $d(\cdot, \cdot)$ is a distance metric, and $f_{\mathrm{sg}}$ denotes the stop-gradient version of the current network. At the early stage of training, $r=0$ is set, and the loss degenerates to standard diffusion model training; then, $r$ is gradually increased to progressively tighten the consistency constraint. In this work, we introduce consistency models into speech synthesis and adopt the ECT method to achieve efficient consistency training.

\section{ECTSpeech}

\begin{figure*}[htbp]
  \centering
  \includegraphics[width=0.96\textwidth]{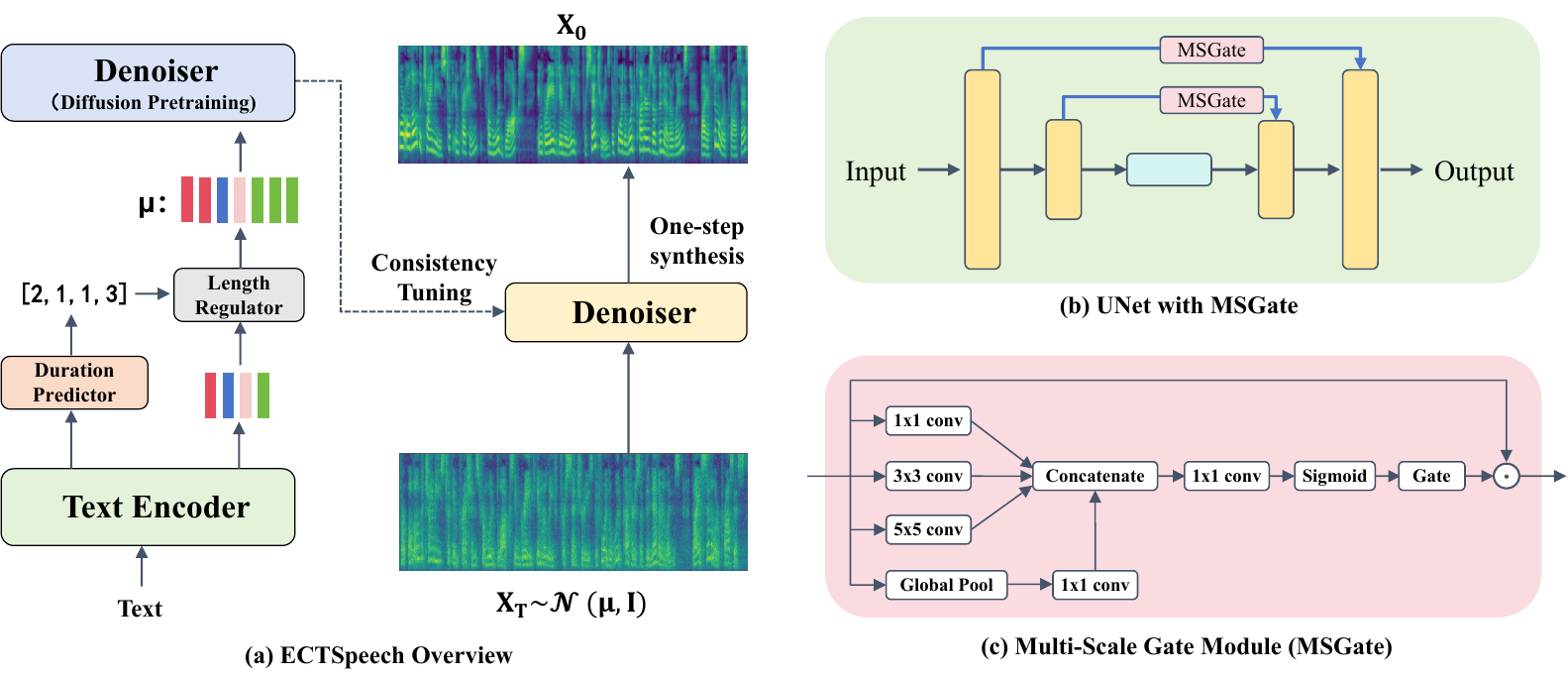}
  \caption{Overview of the proposed ECTSpeech framework. (a) System overview, where the dashed arrow indicates the consistency tuning process during fine-tuning. (b) The UNet decoder with multi-scale gate modules (MSGate) applied to skip connections. (c) Details of the MSGate module, illustrating the multi-branch fusion and gating mechanism.
  }
  \Description{
    Overview of the proposed ECTSpeech framework. (a) System overview, where the dashed arrow indicates the consistency tuning process during fine-tuning. (b) The UNet decoder with multi-scale gate modules (MSGate) applied to skip connections. (c) Details of the MSGate module, illustrating the multi-branch fusion and gating mechanism.
  }
  \label{fig:framework}
\end{figure*}

In this section, we introduce the overall architecture of the ECTSpeech framework. Unlike previous diffusion-based speech synthesis approaches, ECTSpeech employs the ECT strategy to achieve one-step speech generation without distillation. Figure~\ref{fig:framework} presents the system pipeline, and the following subsections describe the MSGate module and the consistency tuning method in detail.

\subsection{Model Overview}

As illustrated in Figure~\ref{fig:framework}(a), ECTSpeech consists of four key components: a text encoder, a duration predictor, a length regulator, and a denoising network. Given a text sequence, the encoder first converts it into phoneme-level hidden representations. The duration predictor then estimates the duration for each phoneme, and the length regulator expands the sequence by repeating and aligning the features according to the predicted durations, producing frame-level features that match the length of the target mel-spectrogram. The frame-level features are linearly projected to generate the prior mel-spectrogram mean $\mu$, which conditions the subsequent U-Net-based denoising network. To improve multi-scale representation learning, MSGate modules are embedded in all U-Net skip connections.

The training of ECTSpeech consists of two stages: \textbf{Diffusion Pretraining} and \textbf{Consistency Tuning}. In the first stage, Diffusion Pretraining, we adopt the Elucidated Diffusion Model (EDM)~\cite{karras2022elucidating} to train the entire model end-to-end for text-to-mel acoustic modeling. Specifically, EDM formulates the process of gradually adding Gaussian noise to the prior mel-spectrogram mean $\mu$ as a probability flow ordinary differential equation (ODE):
\begin{equation}
d\mathbf{x}_t = \frac{\mathbf{x}_t - f_\theta(\mathbf{x}_t, t, \mu)}{t}dt,
\end{equation}
where $\mathbf{x}_t$ denotes the noisy mel-spectrogram at diffusion step $t$, and $f_\theta$ is a parameterized denoising network that predicts the noise-free mel-spectrogram target.

To improve the model's ability to represent features under different noise levels, the denoising network $f_\theta$ adopts a skip-connected structure:
\begin{equation}
f_\theta(\mathbf{x}_t, t, \mu) = c_{\mathrm{skip}}(t)\mathbf{x}_t + c_{\mathrm{out}}(t)F_\theta(\mathbf{x}_t, t, \mu)
\end{equation}
Here, $F_\theta$ denotes the U-Net backbone, while $c_{\mathrm{skip}}(t)$ and $c_{\mathrm{out}}(t)$ are time-dependent functions satisfying $c_{\mathrm{skip}}(\epsilon)=1$ and $c_{\mathrm{out}}(\epsilon)=0$, ensuring model stability and reversibility at the noiseless limit. The training loss in the EDM stage is defined as:
\begin{equation}
\mathcal{L}_{\mathrm{EDM}} = \| f_\theta(\mathbf{x}_t, t, \mu) - \mathbf{x}_0 \|^2
\end{equation}
The EDM-based training enables accurate reconstruction of clean mel-spectrograms, which serves as a strong foundation for consistency tuning.

In the second stage, \textbf{Consistency Tuning}, the text encoder, duration predictor, and length regulator are frozen, and only the denoising network is fine-tuned. This stage introduces the ECT loss to enforce consistency across outputs at different sampling timesteps, enabling one-step generation.

\subsection{Multi-Scale Gate Module}

To enhance multi-scale feature modeling, we incorporate a MSGate module into every skip connection of the U-Net denoising network, as illustrated in Figure~\ref{fig:framework}(b). While standard U-Net architectures in diffusion models typically fuse multi-scale features via simple skip connections, such mechanisms are insufficient to capture the rich and hierarchical characteristics of speech signals—particularly under one-step sampling. MSGate addresses this limitation by adaptively enhancing feature representations through the extraction and fusion of information at multiple scales.

As shown in Figure~\ref{fig:framework}(c), the MSGate module consists of four parallel branches, each designed to capture features at different scales. The $1\times1$ convolution branch focuses on local information along the channel dimension; the $3\times3$ convolution branch captures local spatial details; the $5\times5$ convolution branch extracts broader contextual features with a larger receptive field; and the global pooling branch captures global context via average pooling, followed by a $1\times1$ convolution and upsampling to match the original spatial resolution.

Specifically, the features extracted from all branches are concatenated along the channel dimension and fused via a $1\times1$ convolution, followed by a Sigmoid activation to produce gating weights. These weights are applied element-wise to the input features, enabling adaptive multi-scale fusion. The overall computation of MSGate is defined as:
\begin{equation}
\mathbf{y} = \mathbf{h} \odot \sigma(\mathbf{W}_{\mathrm{fuse}}([\mathbf{h}_{1\times1}; \mathbf{h}_{3\times3}; \mathbf{h}_{5\times5}; \mathbf{h}_{\mathrm{global}}])),
\end{equation}
where $\mathbf{h}$ denotes the input feature, $\mathbf{y}$ the output feature, $\mathbf{W}_{\mathrm{fuse}}$ the fusion weights, $\sigma$ the Sigmoid function, and $\odot$ element-wise multiplication.

\subsection{Easy Consistency Tuning}

After pretraining the diffusion model, we further fine-tune it using the ECT~\cite{geng2025consistency} method to enable high-quality speech generation in a single step. The core idea of ECT is to impose consistency constraints between the model outputs at different noise levels $t$ and $r$, gradually encouraging the model to predict the target data directly at any timestep, and ultimately achieving efficient one-step inference.

Specifically, the ECT stage starts from the pretrained diffusion model and introduces two correlated noise levels, $t$ and $r$, with $r \leq t$. At the early stage of consistency tuning, $r$ is set close to zero. As training progresses, $r$ is gradually increased toward $t$, progressively tightening the consistency constraint. This gradual annealing from $r \approx 0$ to $r \approx t$ enables a smooth transition from a diffusion model to a consistency model, thereby mitigating training instability.

During consistency tuning, a noise sample $\epsilon$ is randomly drawn at each training step to construct two noisy inputs: $\mathbf{x}_t = \mathbf{x}_0 + t\epsilon$ and $\mathbf{x}_r = \mathbf{x}_0 + r\epsilon$. These are then fed into the same denoising network $f_\theta$ to obtain the predictions $f_\theta(\mathbf{x}_t, t, \mu)$ and $f_\theta(\mathbf{x}_r, r, \mu)$, respectively.

We then apply a consistency loss to enforce similarity between the two outputs, thereby enabling consistency training. Specifically, we compute the mean squared error between the outputs, treating the one from the smaller timestep as the target. To ensure training stability, the gradient is stopped on this target output. The loss function is defined as:
\begin{equation}
\mathcal{L}_{\mathrm{ECT}} = \|f_\theta(\mathbf{x}_t, t, \mu) - f_\theta^{\mathrm{sg}}(\mathbf{x}_r, r, \mu)\|^2,
\end{equation}
where $f_\theta^{\mathrm{sg}}(\cdot)$ denotes the stop-gradient operation applied to the network output, treating it as a fixed target during backpropagation. This ensures that the model is optimized to align with the higher-noise output, thereby enabling stable consistency training.

In addition, due to large variations in speech length, the corresponding mel-spectrograms also vary significantly in length. This leads to longer utterances contributing more to the overall loss during training, while shorter ones contribute less, often resulting in degraded quality for short utterances. To address this issue, we introduce a mask-based normalization strategy that balances the loss contribution across samples by focusing on the valid regions. The normalized loss is defined as:
\begin{equation}
\mathcal{L}_{\mathrm{ECT}}^{\mathrm{norm}} = \frac{1}{\sum_{i,j} m_{i,j}}\sum_{i,j}\|f_\theta(\mathbf{x}_t, t, \mu)_{i,j}-f_\theta^{\mathrm{sg}}(\mathbf{x}_r, r, \mu)_{i,j}\|^2 \cdot m_{i,j}
\end{equation}
where $m_{i,j}$ is a binary mask indicating valid regions of the mel-spectrogram (1 for valid frames, 0 for padded frames). This normalization ensures that each sample contributes more equally to the training objective.

\subsection{Training and Loss}

This subsection presents the two-stage training process and loss formulation of ECTSpeech, corresponding to diffusion pretraining and consistency tuning.

In the first stage, we follow the training protocols of prior works such as CoMoSpeech~\cite{comospeech} and Grad-TTS~\cite{Grad-TTS} to pretrain the entire acoustic model. The training objective consists of a duration loss and a prior mel-spectrogram loss. The duration loss is used to optimize the duration predictor and is defined as:
\begin{equation}
\mathcal{L}_{\mathrm{duration}} = \| \log(d_{\mathrm{pred}}) - \log(d_{\mathrm{gt}}) \|^2,
\end{equation}
where $d_{\mathrm{pred}}$ is the predicted phoneme duration, and $d_{\mathrm{gt}}$ is the ground-truth duration label. The prior mel-spectrogram loss supervises the length regulator and the generation of the prior spectrogram:
\begin{equation}
\mathcal{L}_{\mathrm{prior}} = \|\mu - \mathrm{gt}_{\mathrm{mel}}\|^2,
\end{equation}
where $\mu$ is the predicted prior mel-spectrogram, and $\mathrm{gt}_{\mathrm{mel}}$ is the ground-truth mel-spectrogram. For the diffusion acoustic model, we adopt the EDM~\cite{karras2022elucidating}, with the reconstruction loss defined as:
\begin{equation}
\mathcal{L}_{\mathrm{EDM}} = \|f_\theta(\mathbf{x}_t, t, \mu) - \mathbf{x}_0\|^2,
\end{equation}
where $\mathbf{x}_t$ is the noisy mel-spectrogram sample, $f_\theta$ is the parameterized denoising network, and $\mu$ is the conditioning input.

In the second stage, building on the pretrained diffusion model, we freeze the text encoder, duration predictor, and length regulator, and fine-tune only the denoising network. The model is optimized using the ECT loss:
\begin{equation}
\mathcal{L}_{\mathrm{ECT}}^{\mathrm{norm}} = \frac{1}{\sum_{i,j} m_{i,j}}\sum_{i,j}\|f_\theta(\mathbf{x}_t, t, \mu)_{i,j}-f_\theta^{\mathrm{sg}}(\mathbf{x}_r, r, \mu)_{i,j}\|^2 \cdot m_{i,j}
\end{equation}

During consistency tuning, we maintain an exponential moving average (EMA) version of the denoising network. The EMA parameters are continuously updated during training and are used exclusively for speech generation during inference, without participating in backpropagation.

\begin{table*}[htbp]
  \centering
  \caption{Evaluation results on LJSpeech for TTS. }
  \label{tab:main_results}
  \begin{tabular}{lccccc}
    \toprule
    Method & NFE & FAD ($\downarrow$) & FD ($\downarrow$) & RTF ($\downarrow$) & MOS ($\uparrow$) \\
    \midrule
    GT & \textendash & \textendash & \textendash & \textendash & $4.68 \pm 0.10$ \\
    GT (Mel+Voc) & \textendash & 0.2709 & 0.3858 & \textendash & $4.55 \pm 0.09$ \\
    \midrule
    FastSpeech2 & 1 & 1.6125 & 10.8170 & \textbf{0.0027} & $3.94 \pm 0.10$ \\
    DiffGAN-TTS & 4 & 1.7281 & 6.8291 & 0.0087 & $3.81 \pm 0.07$ \\
    Grad-TTS & 50 & 1.0792 & 2.4317 & 0.1405 & $4.28 \pm 0.11$ \\
    CoMoSpeech (teacher) & 50 & 0.6025 & 0.6819 & 0.1538 & $4.33 \pm 0.10$ \\
    CoMoSpeech & 1 & 0.6297 & 0.6783 & 0.0070 & $4.19 \pm 0.09$ \\
    \midrule
    ECTSpeech (pre-trained) & 50 & 0.5738 & \textbf{0.6328} & 0.1589 & \textbf{4.35 $\pm$ 0.09} \\
    ECTSpeech & 1 & \textbf{0.5246} & 0.6976 & 0.0077 & $4.16 \pm 0.08$ \\
    \bottomrule
  \end{tabular}
\end{table*}

\section{Experiments}

\subsection{Data and Preprocessing}

We conduct experiments on the public LJSpeech dataset~\cite{ljspeech}, which contains 13,100 English audio clips (approximately 24 hours in total) recorded by a single female speaker at a sampling rate of 22.05 kHz. The dataset is split into 93.3\% for training (12,228 samples), 2.7\% for validation (349 samples), and 4.0\% for testing (523 samples). All audio samples are converted into 80-dimensional mel-spectrograms with a frame length of 1024 and a hop size of 256.

\subsection{Model Setup}

We adopt the same text encoder and duration predictor architecture as CoMoSpeech. The encoder consists of six Feed-Forward Transformer (FFT) blocks~\cite{fastspeech}, each with a hidden size of 192. The duration predictor is composed of two convolutional layers. The pretrained diffusion model is trained on a single NVIDIA RTX 3090 GPU for 1.7 million steps with a batch size of 16, using the Adam optimizer~\cite{adam} and a learning rate of 1e-4. The ECTSpeech consistency tuning stage is also conducted on a single RTX 3090 GPU for 170k steps with the same batch size and optimizer, and a reduced learning rate of 1e-5. During consistency tuning, only the denoising network is updated while all other modules are frozen.

We compare our model against GT, GT (Mel+Voc), FastSpeech2~\cite{fastspeech2}, DiffGAN-TTS~\cite{diffgan}, Grad-TTS~\cite{Grad-TTS}, and CoMoSpeech~\cite{comospeech}. GT refers to original recordings, and GT (Mel+Voc) uses ground-truth mel-spectrograms and generates speech through vocoder-based reconstruction. FastSpeech2 is a fast, non-autoregressive TTS model; DiffGAN-TTS employs adversarial training for efficient mel generation; Grad-TTS applies diffusion modeling with probability flow sampling; CoMoSpeech uses consistency distillation for one-step synthesis. All models use a pretrained HiFi-GAN~\cite{hifi-gan} vocoder to convert mel-spectrograms into waveforms.

\subsection{Evaluation Metrics}

We conduct both subjective and objective evaluations of sample quality and model inference efficiency, including Fréchet Distance (FD), Fréchet Audio Distance (FAD), Mean Opinion Score (MOS), Real-Time Factor (RTF), and Number of Function Evaluations (NFE). FD and FAD are adapted from the Fréchet Inception Distance (FID)~\cite{heusel2017gans} commonly used in image generation, and measure the distributional discrepancy between generated and real samples. Following the implementation in~\cite{audioldm}, FD features are extracted using a PANNs classifier, while FAD features are extracted using a VGGish classifier. Lower scores on both metrics indicate higher similarity to real speech.

MOS~\cite{mos} is rated on a scale from 1 to 5, assessing naturalness and intelligibility. For each system, 10 utterances are randomly selected and rated by 10 human evaluators. Higher scores indicate better perceptual quality. RTF reflects the actual time required to generate one second of audio and serves as a measure of real-time performance. NFE (Number of Function Evaluations) refers to the total number of denoising network calls during generation and is used to quantify computational cost.

\begin{figure}[htbp]
  \centering
  \includegraphics[width=1\linewidth]{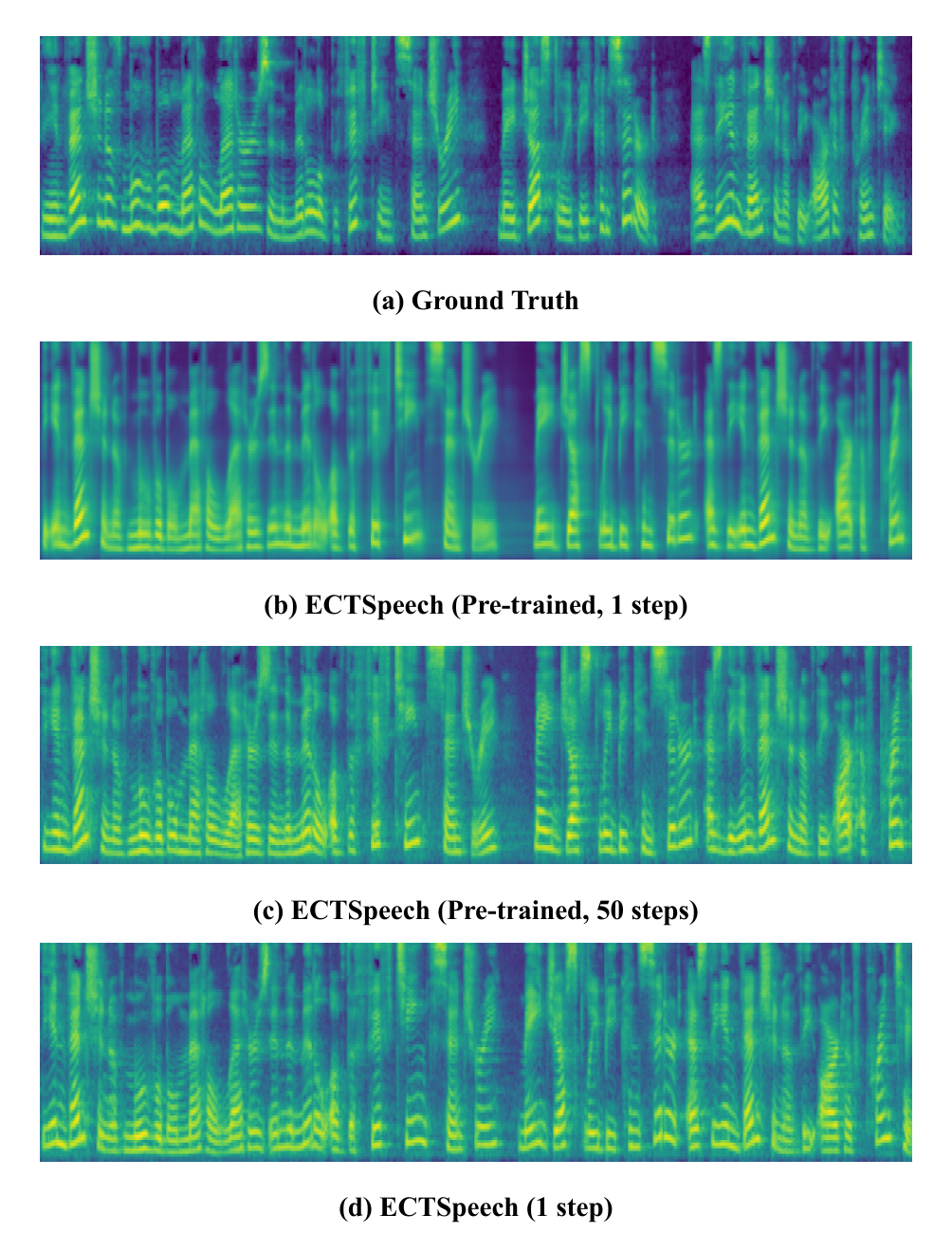}
  \caption{
    Mel-spectrograms for qualitative comparison: 
    (a) Ground Truth; 
    (b) ECTSpeech (Pre-trained, 1 step); 
    (c) ECTSpeech (Pre-trained, 50 steps); 
    (d) ECTSpeech (1 step).
  }
  \label{fig:mel-qualitative}
\end{figure}

\subsection{Audio Performance}

The TTS evaluation results are summarized in Table~\ref{tab:main_results}. In terms of speech quality, our ECTSpeech (pre-trained) achieves the highest MOS and the best FD score, demonstrating strong capability in modeling the distribution of natural speech. Notably, ECTSpeech reaches comparable single-step quality to CoMoSpeech with only 10\% of its training iterations. Moreover, it significantly outperforms FastSpeech2 and DiffGAN-TTS—two fast-inference baselines—in both perceptual quality and distributional metrics. ECTSpeech also achieves the best FAD score among all compared methods, further validating the effectiveness of diffusion pretraining and the consistency tuning strategy.

We further perform a qualitative analysis of the generated mel-spectrograms, as shown in Figure~\ref{fig:mel-qualitative}. The results indicate that ECTSpeech (pre-trained) with 50-step sampling captures rich spectral details, while the single-step output exhibits some degree of blurriness. After consistency tuning, ECTSpeech (1 step) produces significantly sharper mel-spectrograms, closely matching both the ground-truth and multi-step results in structure and detail. This demonstrates that consistency tuning substantially improves the naturalness and fidelity of single-step speech generation.

\begin{table}[h]
  \centering
  \caption{Ablation Study on LJSpeech. All results are based on 1-step inference.}
  \label{tab:ablation}
  \begin{tabular}{lccc}
    \toprule
    Method & FAD ($\downarrow$) & FD ($\downarrow$) & MOS ($\uparrow$) \\
    \midrule
    ECTSpeech & 0.5246 & 0.6976 & 4.16 $\pm$ 0.08 \\
    \midrule
    w/o MSGate & 0.6621 & 0.7453 & 4.09 $\pm$ 0.09 \\
    w/o Masked Normalization & 0.6947 & 0.7031 & 4.11 $\pm$ 0.10 \\
    w/o Consistency Tuning & 1.6863 & 7.9731 & 3.41 $\pm$ 0.12 \\
    \bottomrule
  \end{tabular}
\end{table}

\subsection{Ablation Studies}

To further evaluate the contribution of each component to the overall performance of ECTSpeech, we conduct ablation studies on the LJSpeech test set. All experiments are performed under the single-step sampling setting, and the results are presented in Table~\ref{tab:ablation}.

Removing the MSGate module leads to an increase in FAD and FD to 0.6621 and 0.7453, respectively, and a drop in MOS to 4.09, indicating that the multi-scale gating mechanism positively contributes to both perceptual quality and distributional alignment. \textit{Masked Normalization} refers to normalizing the consistency loss by the number of valid frames across samples of varying lengths. Disabling this strategy also results in degraded performance, with higher FAD and FD scores and lower MOS, suggesting its effectiveness in improving overall synthesis quality. Notably, without consistency tuning (w/o Consistency Tuning), i.e., directly applying one-step sampling to the pretrained model, performance deteriorates significantly across all metrics—FD rises to 7.9731, FAD to 1.6863, and MOS drops to 3.41. These results highlight that consistency tuning is essential for achieving high-quality single-step speech synthesis.

\section{Conclusion}

In this paper, we propose ECTSpeech, an efficient speech synthesis framework based on Easy Consistency Tuning. Built upon a pretrained diffusion acoustic model, ECTSpeech enables high-quality one-step generation through consistency constraints, without requiring an additional teacher model. The proposed MSGate module effectively enhances the model’s ability to adaptively fuse multi-scale features. Experimental results demonstrate that ECTSpeech outperforms or matches state-of-the-art methods in terms of training cost, inference efficiency, and speech quality. In future work, we plan to explore more lightweight designs and extend the framework to multi-speaker and emotional speech synthesis scenarios.


%%\begin{acks}
%%To Robert, for the bagels and explaining CMYK and color spaces.
%%\end{acks}
\section{Acknowledgments}
This research was financially supported by the National Natural Science Foundation of China (Grant No. 62463029).
\bibliographystyle{ACM-Reference-Format}
\bibliography{sample-base}

\end{document}